\documentclass[twocolumn]{aastex631}

\usepackage{url}
\usepackage{hyperref}

\hypersetup{
     colorlinks=true, 
     linkcolor=red,  
     filecolor=magenta,   
     citecolor=blue,   
     urlcolor=cyan,  
     menucolor=black
     }

\shorttitle{Dynamical state of SMC clusters}
\shortauthors{Dresbach et al.}
\graphicspath{{./}{figures/}}

\begin{document}

\title{Blue Stragglers as tracers of the dynamical state of two clusters in the Small Magellanic Cloud: NGC\,339 and NGC\,419}



\author{F. Dresbach}
\affiliation{Dept. of Physics and Astronomy, University of Bologna, Via Gobetti 93/2, Bologna, Italy}

\author{D. Massari}
\affiliation{INAF - Osservatorio di Astrofisica e Scienza dello Spazio di Bologna, Via Gobetti 93/3, I-40129 Bologna, Italy}
\affiliation{Univerity of Groningen, Kapteyn Astronomical Institute, NL-9747 AD Groningen, The Netherlands}

\author{B. Lanzoni}
\affiliation{Dept. of Physics and Astronomy, University of Bologna, Via Gobetti 93/2, Bologna, Italy}
\affiliation{INAF - Osservatorio di Astrofisica e Scienza dello Spazio di Bologna, Via Gobetti 93/3, I-40129 Bologna, Italy}

\author{F. R. Ferraro}
\affiliation{Dept. of Physics and Astronomy, University of Bologna, Via Gobetti 93/2, Bologna, Italy}
\affiliation{INAF - Osservatorio di Astrofisica e Scienza dello Spazio di Bologna, Via Gobetti 93/3, I-40129 Bologna, Italy}

\author{E. Dalessandro}
\affiliation{INAF - Osservatorio di Astrofisica e Scienza dello Spazio di Bologna, Via Gobetti 93/3, I-40129 Bologna, Italy}

\author{S. Raso}
\affiliation{INAF - Osservatorio di Astrofisica e Scienza dello Spazio di Bologna, Via Gobetti 93/3, I-40129 Bologna, Italy}

\author{A. Bellini}
\affiliation{Space Telescope Science Institute, 3700 San Martin Drive, Baltimore, MD 21218, USA}

\author{M. Libralato}
\affiliation{AURA for the European Space Agency (ESA), ESA Office, Space Telescope Science Institute, 3700 San Martin Drive, \\ Baltimore, MD 21218, USA}

\begin{abstract}
\noindent The level of central segregation of Blue Straggler stars proved to be an excellent tracer of the dynamical evolution of old star clusters (the so-called “dynamical clock"), both in the Milky Way and in the Large Magellanic Cloud. The  $A^+$ parameter, used to measure the Blue Stragglers degree of segregation, has in fact been found to strongly correlate with the parent cluster central relaxation time. 
Here we studied the Blue-Straggler population of two young stellar systems in the Small Magellanic Cloud, namely NGC\,339 (which is 6 Gyr old) and NGC\,419 (with an age of only 1.5 Gyr), in order to study their dynamical state. Thanks to multi-epoch, high angular resolution Hubble Space Telescope observations available for both clusters, we took advantage of the stellar proper motions measured in the regions of the two systems and we selected a population of likely cluster members, removing the strong contamination from Small Magellanic Cloud stars. This enabled us to study, with unprecedented accuracy, the radial distribution of Blue Stragglers in these two extragalactic clusters and to measure their dynamical age. As expected for such young clusters, we found that both systems are poorly evolved from the dynamical point of view, also fully confirming that the $A^+$ parameter is a sensitive “clock hand” even in the dynamically-young regime.

\end{abstract}


\section{Introduction} \label{sec:intro}
Globular Clusters (GCs) are dynamically active systems, where the frequent gravitational interactions among stars can drastically alter the internal structure. The time-scale of a cluster dynamical evolution depends on a variety of its properties, both internal (total mass, density, fraction of binaries, etc.) and external (tidal interaction in the host galaxy, local density). Determining the evolutionary state of GCs is therefore a complex task. Dynamical interactions also lead to the formation of exotic objects like Blue Straggler stars (BSSs). These peculiar stars are located on the extension of the main sequence (MS), in a color-magnitude diagram (CMD), in the region that is hotter and bluer than the turn-off point (TO) (\citealt{Sandage1953}, \citealt{Ferraro1992}, \citeyear{Ferraro1993}, \citeyear{ferraro1997}). BSSs are thought to be the result of two mass-enhancement processes, either mass-transfer in binary systems \citep{1964McCrea} or stellar mergers resulting from direct collisions (\citealt{1976hills}). Since BSSs are more massive than the average (in Galactic GCs they have masses of $\sim 1.2~ \rm{M_{\odot}}$, while the average stellar mass is $\left \langle m \right \rangle \sim 0.3 ~\rm{M_{\odot}}$; \citealt{shara1997}, \citealt{2014fiore}, \citealt{raso2019}), their level of central segregation is an indicator of the host system dynamical age. This concept was for the first time demonstrated  from the analysis of the BSS radial distribution (yielding to the definition of the so-called dynamical clock (\citealt{Ferraro2012}) and then further defined by using the $A^+$ parameter (\citealt{Lanzoni2016}, \citealt{ferraro2018}). This parameter is defined (\citealt{alessandrini2016}) as the area enclosed between the cumulative radial distribution of BSSs and that of a reference (and lighter) population. As such, its value increases as the dynamical evolution of the cluster proceeds and makes heavier stars sink in more rapidly than the less massive ones (\citealt{alessandrini2016}, \citealt{Lanzoni2016}). Indeed, this parameter shows a tight correlation with structural/dynamical parameters, such as central relaxation time and core radius, which trace the cluster dynamical ageing (\citealt{ferraro2018}). The same correlations were then confirmed in a few old extra-Galactic clusters located in the Large Magellanic Cloud \citep{FERRARO2019}. However, this kind of analysis has not yet been made for young star clusters. To this end, the Small Magellanic Cloud (SMC) represents an ideal environment since it is populated by several young (t $<$ 2 Gyr) and intermediate age clusters ($\rm{t}=$ 3 -- 7 Gyr). In this paper we analyze the Blue Straggler population of two star clusters, NGC\,339 and NGC\,419, located in this galaxy and with an age of 6 Gyr and 1.5 Gyr, respectively (\citealt{glatt2009}). The study of their BSS populations has been hindered by the large contamination from SMC field stars that occupy the same region of the CMD. Traditional statistical techniques used to decontaminate from non-cluster members (e.g. \citealt{CABRERA}, \citealt{dalessandro2019}) may not work on a star-by-star basis, and thus could artificially alter the spatial distribution of the decontaminated sample. For this reason, a kinematic decontamination acting on each individual star is required. Thanks to the availability of multi-epoch observations obtained for both clusters with the \textit{Hubble Space Telescope} (\textit{HST}), from which relative proper motions (PMs) of individual stars have been measured \citep{massari2021}, we have performed a kinematic study aimed at selecting BSS members of each cluster. Our final goal is to determine the dynamical age of these clusters and verify whether the BSS population can act as a dynamical indicator even in these younger systems.

The paper is organised as follows. In Section \ref{sec:data} we present the dataset and the PM measurements. In Section \ref{sec:selCyn} we describe the decontamination procedure and the cluster membership analysis. In Section \ref{sec:dyn} we study the BSS population, determining the $A^{+}$ parameter and the dynamical state of the clusters, and we discuss the results. The conclusions are provided in Section \ref{sec:conclusio}.

\section{Data analysis} \label{sec:data}

The data available for the two clusters have been obtained from observations made with \textit{HST}. Images in the F336W and F438W filters were acquired using the Ultraviolet-Visible Channel (UVIS) of the Wide Field Camera 3 (WFC3), while the Wide Field Channel (WFC) of the Advanced Camera for Surveys (ACS) was used to acquire the images in the F555W and F814W filters. The list of observations for NGC\,339 and NGC\,419 is given in Table \ref{tab:images339} and Table \ref{tab:images419}, respectively. These multi-epoch observations provide long temporal baselines for the PM measurements, of 10.75 yr for NGC\,339 and 12.67 yr for NGC\,419.

\begin{table*}[h]
\begin{center}
\begin{tabular}{ c c c c c l }
 \hline\hline
 Program ID & PI & Epoch & Camera & Filter & Exposures \\
  &&(yyyy/mm)&&&N $\times$ $t_{\rm exp}$\\
 \hline
 GO-10396&J. Gallagher &2005/11  &ACS/WFC&F555W& 2 $\times$ 20 s\\
 &&&&&4 $\times$ 496 s\\
 &&&&F814W&2 $\times$ 10 s\\
 &&&&& 4 $\times$ 474 s\\
 \hline
 GO-14069 &N. Bastian &2016/08 &WFC3/UVIS &F336W & 2 $\times$ 1200 s\\
 &&&&& 1 $\times$ 700 s\\
 &&&&F438W &1 $\times$ 120 s\\
 &&&&&1 $\times$ 180 s\\
 &&&&&1 $\times$ 560 s\\
 &&&&&1 $\times$ 660 s\\
 \hline
\end{tabular}
\caption{List of the observations of NGC\,339 obtained with \textit{HST} used in this work.}
\label{tab:images339}
\end{center}
\renewcommand{\arraystretch}{1.0}
\end{table*}

\begin{table*}
\begin{center}
\begin{tabular}{ c c c c c l }
 \hline\hline
 Program ID & PI & Epoch & Camera & Filter & Exposures \\
  &&(yyyy/mm)&&&N $\times$ $t_{\rm exp}$\\
 \hline
 GO-10396&J. Gallagher&2006/01&ACS/WFC&F555W&1 $\times$ 20 s\\
 &&&&F814W&2 $\times$ 10 s\\
 &&&&&4 $\times$ 474 s\\
 &&2006/07&ACS/WFC&F555W&2 $\times$ 20 s\\
 &&&&&4 $\times$ 496 s\\
 &&&&F814W&2 $\times$ 10 s\\
 &&&&&4 $\times$ 474 s\\
 \hline
 GO-12257&L. Girardi&2011/08&WFC3/UVIS&F336W&1 $\times$ 400 s\\
 &&&&&1 $\times$ 690 s\\
 &&&&&2 $\times$ 700 s\\
 &&&&&1 $\times$ 740 s\\
 \hline
 GO-14069&N. Bastian&2016/08 &WFC3/UVIS& F438W&1 $\times$ 70 s\\
 &&&&&1 $\times$ 150 s\\
 &&&&&1 $\times$ 350 s\\
 &&&&&1 $\times$ 550 s\\
 \hline
 GO-15061 &N. Bastian &2018/09& WFC3/UVIS &F336W &2 $\times$ 1395 s\\
 &&&&&1 $\times$ 3036 s\\
 &&&& F438W &2 $\times$ 1454 s\\
 \hline
\end{tabular}
\caption{List of the observations of NGC\,419 obtained with \textit{HST} used in this work.}
\label{tab:images419}
\end{center}

\renewcommand{\arraystretch}{1.0}
\end{table*}

The photometric reduction is performed following the prescriptions given in \cite{bellini2017}, \cite{bellini2018} and summarised in the following. We analyzed \texttt{\_flc} images that were also corrected for charge transfer efficiency (CTE, \citealt{2010andersonBedin}). Briefly, we first performed a single-pass photometry to measure bright stars in each exposure, without subtracting the neighbouring sources. Then, after correcting the instrumental positions for geometric distortions with the solutions provided in \cite{andersonking2006}, \cite{belliniAndersonBedin2011}, we performed a multi-pass photometry with the {\tt KS2} program. This simultaneously analyzes all the available images in all the filters, by combining the results of the previous step and also performing neighbour subtraction. Instrumental magnitudes were calibrated into the \texttt{VEGAMAG} photometric system, following the example of \cite{Bellini2017a} and \cite{Raso2020}.

The relative PM measurements of NGC~419 used in this work are those determined by \cite{massari2021}. We refer to this paper for the details of the PM measurement procedure, which is based on the techniques developed by \cite{Bellini20142014}, later improved by \cite{bellini2018} and \cite{libralato2018}. Briefly, the procedure to measure the relative PMs is iterative. At each step, single-exposure star positions are transformed on to an epoch-matched reference frame (based on the \textit{Gaia} Data Release 2 catalog) by means of a six-parameter linear transformation. Only bright and unsaturated cluster stars are used at this stage. At the first iteration, PMs are assumed to be zero, and cluster members are solely defined based on their position on a CMD. For each star, locally-transformed positions as a function of epoch are linear least-squares fitted. Local transformations help in mitigating small-scale systematic effects. The slope of these fits are direct estimates of the stars PMs. We apply sophisticated data rejection and sigma-clipping at each iteration. Convergence is reached when the predicted difference between the master-frame positions at the reference epoch from one step to the next is negligible. After convergence, spatially variable and colour-dependent systematic effects are corrected as described in \cite{Bellini20142014}. The relative PMs for NGC~339 were measured following the same procedures.

\subsection{Selection criteria}\label{subsec:criteria}
\begin{figure}[h!]
\includegraphics[scale=0.45]{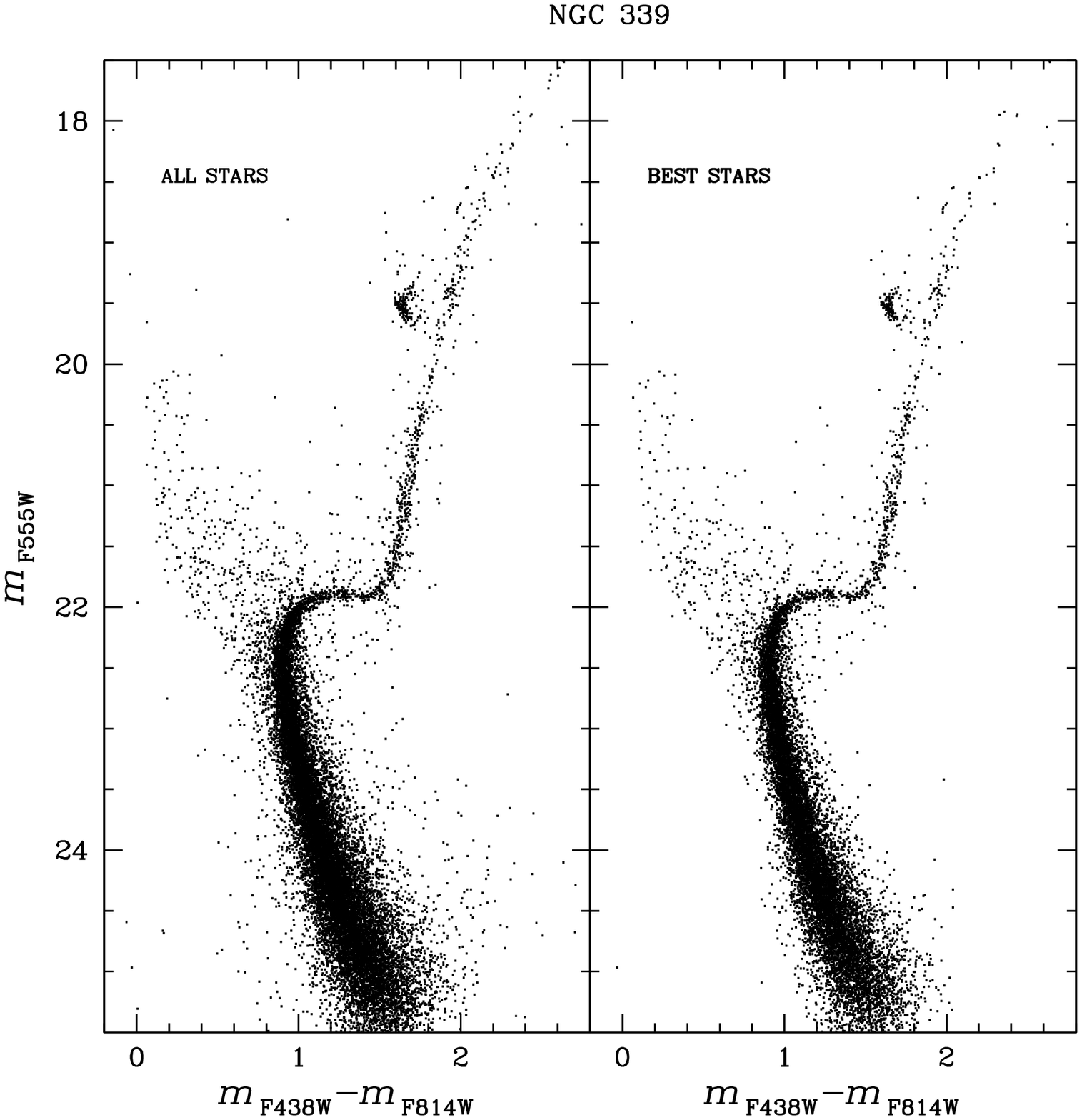}
\includegraphics[scale=0.45]{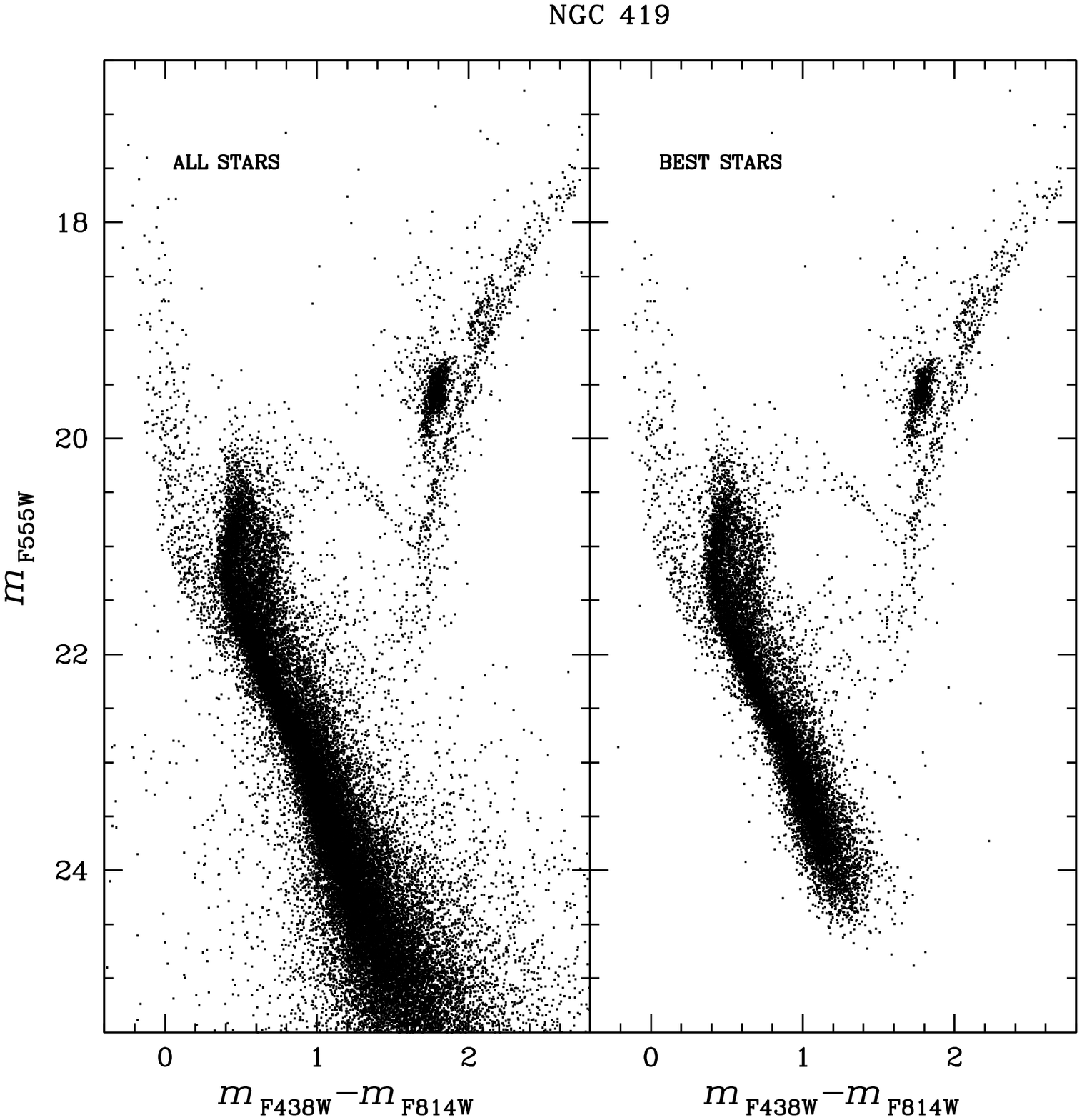}
\caption{CMDs of NGC\,419 (top) and NGC\,339 (bottom) built from the catalogues improved by the quality selection discussed in Section \ref{subsec:criteria} (right), compared with the complete CMD catalogues (left).}
\label{fig:cleanCMD}
\end{figure}

To remove stars that are poorly measured from our catalogues, we applied several photometric and astrometric quality criteria, following the guidelines provided by \cite{libralato2019}. Stars are selected as follows: (i) an iterative 3$\sigma$-clipping procedure is applied to the photometric error of all the 4 filters, removing stars with the higher uncertainties: the procedure is stopped after 5 iterations; (ii) the same procedure is applied to PM errors;
(iii) we selected stars for which the reduced $\chi^2$ of the PM fit is smaller than two in both PM components;
(iv) we applied an iterative 3$\sigma$-clipping selection around the mean value of the \texttt{RADXS}\footnote{A measurement of how much flux there is in the pixels just outside of the core, in excess of the prediction from the PSF; it is positive if the object is
broader than the PSF, and negative if it is sharper (\citealt{bedin}).} parameter;
(v) we eliminated sources having a ratio $\rm{N_g/N_u}<0.8$, where $\rm{N_u}$ is the total number of detections for each source and $\rm{N_g}$ is the number of detections considered of sufficiently good quality by the adopted program. The CMDs of the two clusters resulting from such a selection are shown in Figure \ref{fig:cleanCMD} and compared with the CMDs built from the complete catalogues. The high quality of the available data allows us to recognize sources 3 -- 4 magnitudes below the TO point, even in very crowded regions, which is unfeasible with ground-based observations. It is also possible to clearly identify the different evolutionary sequences, both of the clusters and of the Cloud. Specifically, in NGC\,419 we notice the extended main sequence turn off (eMSTO) and two distinct sub giant branches, one of the cluster (the brighter one) and one (the fainter) populated by stars of the SMC ( see \citealt{massari2021}).

\section{Cluster Membership}\label{sec:selCyn}
\begin{figure*}[h!]
\centering
\includegraphics[scale=0.37]{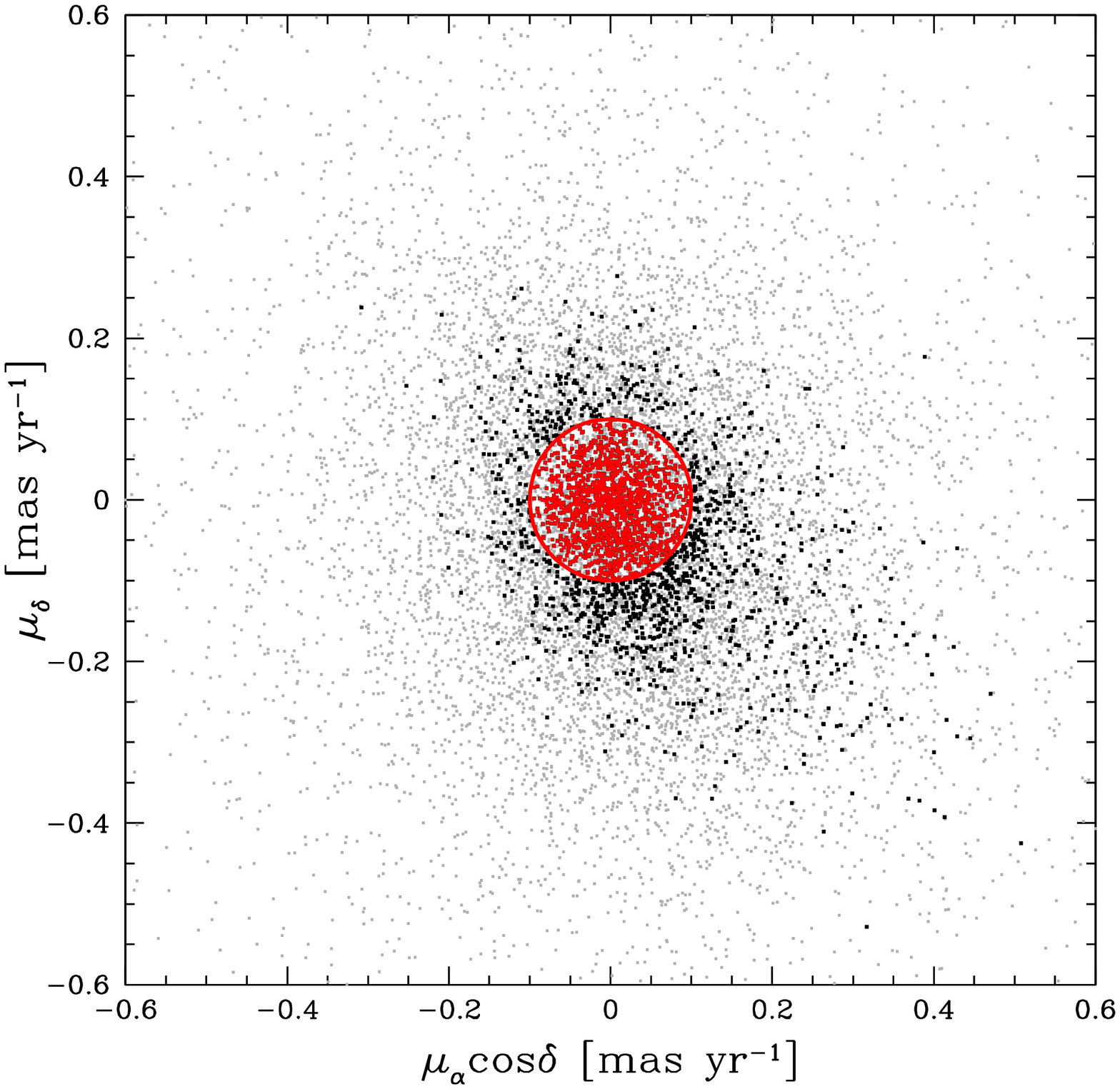}
\includegraphics[scale=0.37]{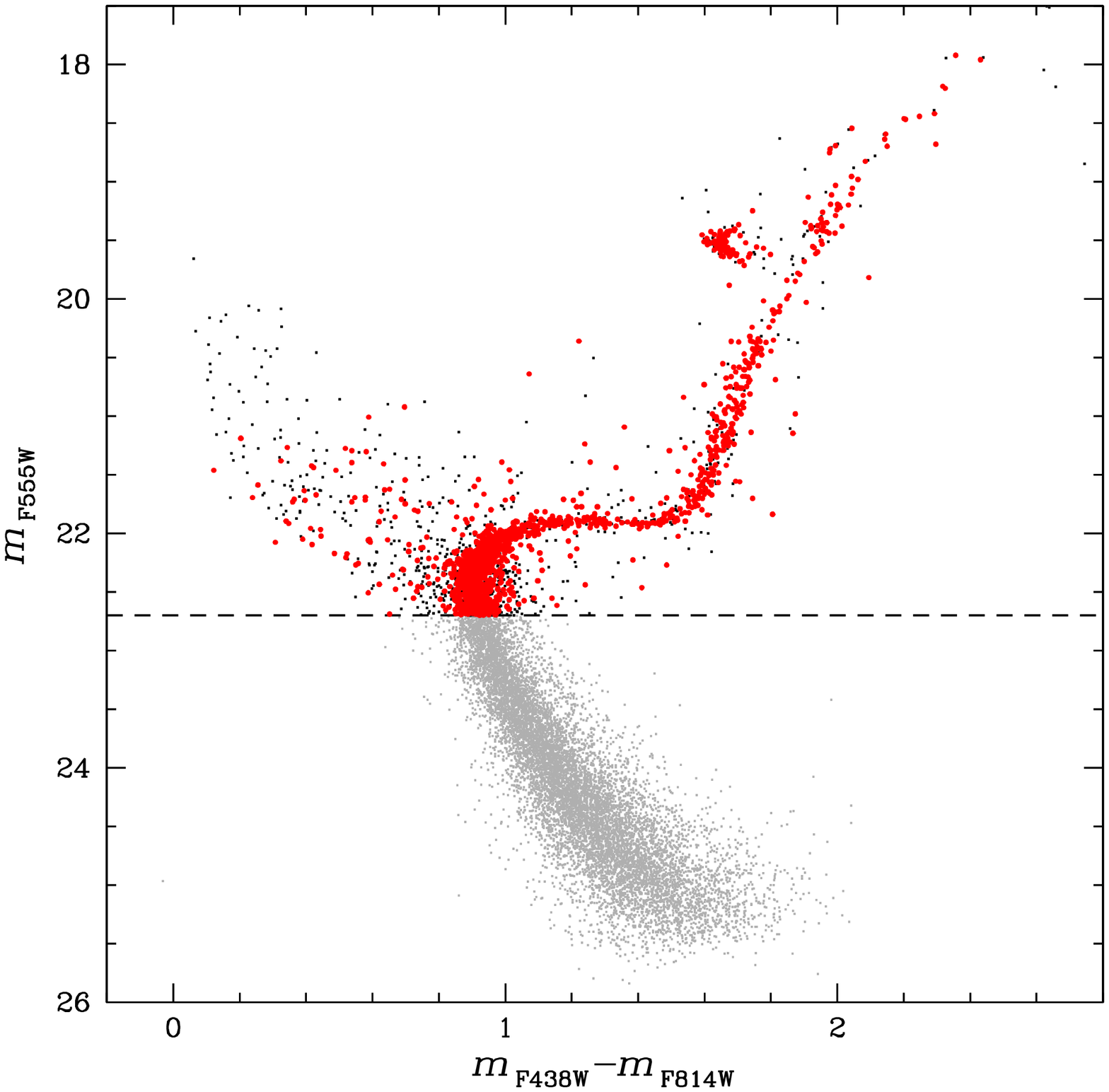}
\caption{VPD (left) and CMD (right) of NGC\,339 for the best stars plotted on the right in Fig. \ref{fig:cleanCMD}. These best stars are plotted here in grey, the stars with $m_{\rm F555W}<22.7$ are shown as black dots and those selected as cluster members are highlighted in red. }
\label{fig:2sigma339}
\end{figure*}

\begin{figure*}[h!]
\centering
\includegraphics[scale=0.37]{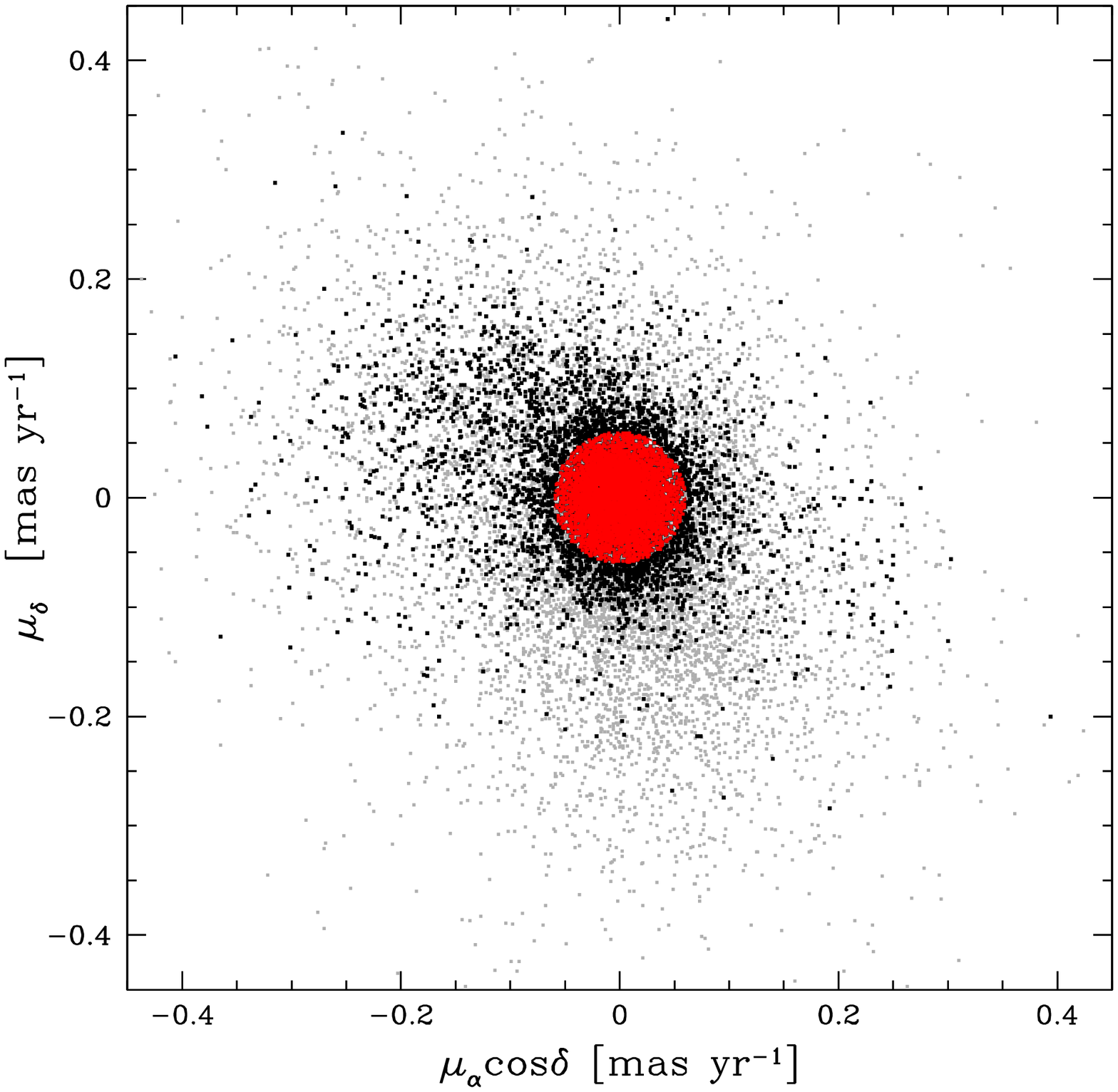}
\includegraphics[scale=0.37]{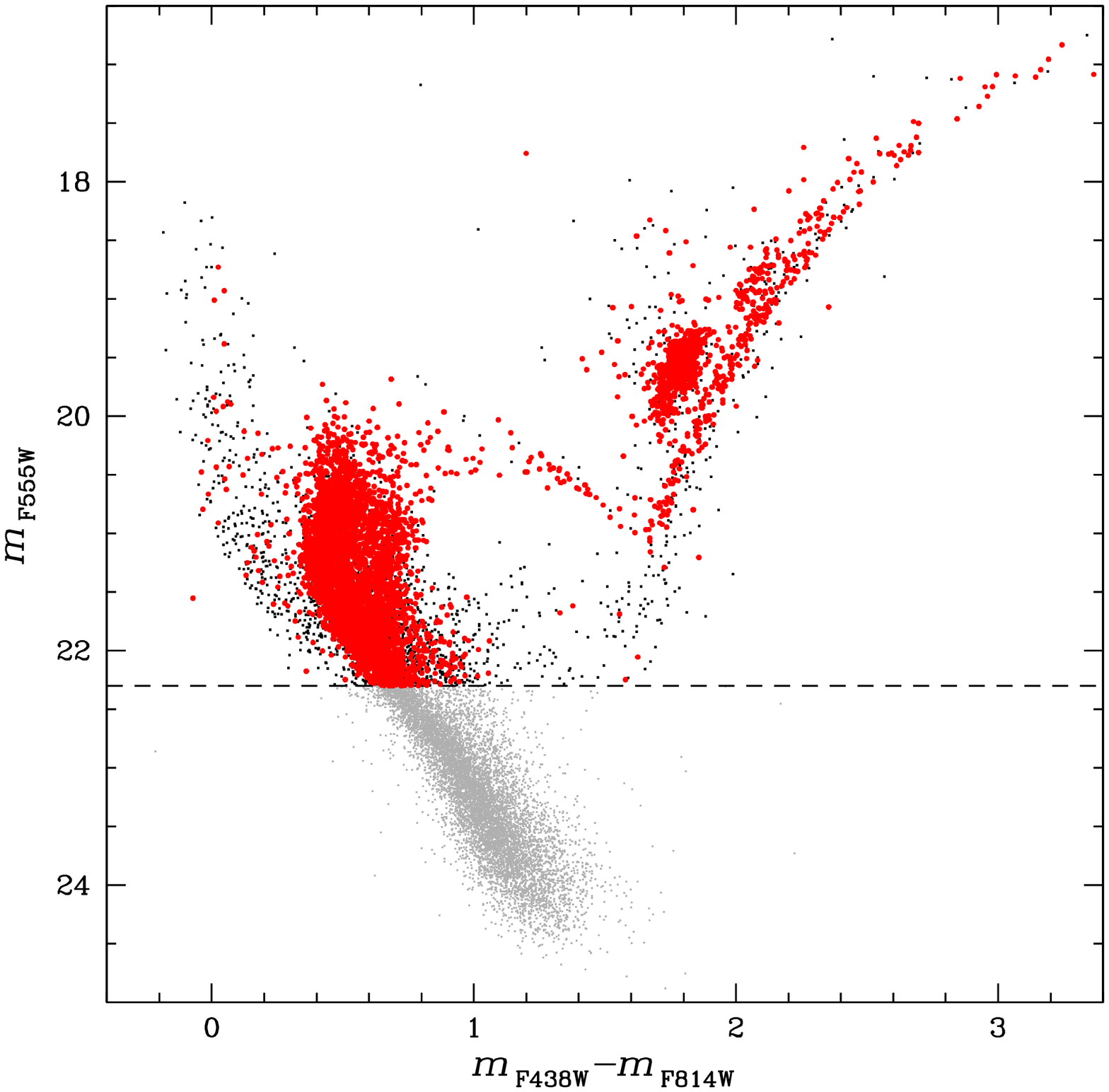}
\caption{Similar to Figure \ref{fig:2sigma339}, but for NGC\,419.}
\label{fig:2sigma419}
\end{figure*}

To study the BSS population of these clusters, a crucial step is to remove the contamination from field stars (SMC + Milky Way). To do so, we took advantage of the kinematical measurements. First of all, we considered only bright stars, for which photometric and PM uncertainties are smaller: those with $m_{\rm F555W} < 22.7$ in NGC\,339, and those with $m_{\rm F555W} < 22.3$ in NGC\,419. The PM distributions of the selected stars are displayed in the vector point diagrams (VPDs; black and red dots in left panels of Figures \ref{fig:2sigma339} and \ref{fig:2sigma419}). To select cluster-members, we thus considered only the stars centered around (0,0) $\rm{mas \ yr^{-1}}$ in the VPD and included within a radius twice the expected total dispersion of cluster members $\sigma$. This is computed as the sum in quadrature of two independent terms. The first is the intrinsic velocity dispersion of the cluster, $\sigma_{\rm disp}$, while the second is the error associated to the PM measurement, $\sigma_{\rm PM}$. Under the assumption of isotropy, the expected dispersion along the two PM components coincide with the dispersion along the line-of-sight, which has been measured to be $\sigma_{\rm los}=2.11 \ \rm{km \ s^{-1}}$ for NGC\,339 \citep{McLaughlin2005} and $\sigma_{\rm los}=2.44 \ \rm{km \ s^{-1}}$ for NGC\,419 \citep{song2019}. At the distance of the SMC ($d_{\rm SMC}\sim 60$ kpc, \citealt{cioni2000}), these both correspond to $\sim 0.01\ \rm{mas \ yr^{-1}}$. As for the PM uncertainties $\sigma_{\rm PM}$, these depend on signal-to noise ratio of each source, hence they vary with the magnitudes. For this reason, we decided to select the average value of the PM error at the faintest magnitude at which the BSS can be detected in a CMD. The $\sigma_{\rm PM}$ values considered this way are: $\sigma_{\rm PM}=0.05\ \rm{mas \ yr^{-1}}$ (at $m_{\mathrm{F555W}}=22.7$) for NGC\,339 and $\sigma_{\rm PM}=0.03\ \rm{mas \ yr^{-1}}$ (at $m_{\mathrm{F555W}}=22.3$) for NGC\,419. By adding in quadrature the two terms ($\sigma=\sqrt{\sigma_{\rm los}^2+\sigma_{\rm PM}^2}$) we obtain a total dispersion for the two clusters of $\sigma_{339}=0.05\ \rm{mas \ yr^{-1}}$ and $\sigma_{419}=0.03\ \rm{mas \ yr^{-1}}$. 
The location in the VPD of the likely members so selected is shown in Figures \ref{fig:2sigma339} and \ref{fig:2sigma419} (red dots). On the CMDs we can see that, thanks to the kinematic selection, we eliminated most of the interlopes located above the MSTO, where BSSs and field stars overlap, thus remarking why the kinematic decontamination is so crucial for our analysis.

A kinematic selection at 3 (instead of 2) $\sigma$ would have included a  larger number of members\footnote{Considering a Gaussian distribution, a $3\sigma$ selection should include $\sim 99.7\%$ of cluster stars, a $2\sigma$ selection only $\sim 95\%$.}, at the cost of allowing for a more significant contamination. The adopted 2$\sigma$ selection ensures the best relative contamination, which we statistically estimated using a Monte Carlo approach.
Starting with NGC\,339, we first fitted the observed distributions of the two PM components of non-member stars with Gaussian functions. We then randomly generated 1000 such Gaussian distributions by keeping mean, sigma and number of elements fixed, each time counting the number of randomly simulated field stars falling within the selection range of cluster members. The mean relative contamination over these 1000 realisations turned out to be $\sim5\%$.
When only considering stars within one half-mass radius\footnote{The value of the half-mass radius is $r_{\rm h}=68\farcs78$ and it was computed from the known value of the projected half-light radius, as described in appendix \ref{app:RH}.} ($r_{\rm h}$, which is the area within which the analysis of the BSSs will be performed, see \citealt{Lanzoni2016}, \citealt{ferraro2018}), the mean contamination further decreases to $\sim4\%$. Note that these numbers refer to the global contamination of field stars compared to cluster members. Another useful information is the relative contamination obtained by restricting the analysis to the CMD region where BSSs are located. To estimate this, we selected a sub-sample of field stars with the following criteria: magnitude $m_{\mathrm{F555W}}<22.7$, color $m_{\mathrm{F438W}}-m_{\mathrm{F814W}}<0.3$ and distance from the cluster centre $r>r_{\rm c}=29\arcsec$ (where $r_{\rm c}$ is the core radius obtained by \citealt{glatt2009}). Then, we repeated the Monte Carlo procedure for the PM distribution of this sub-sample and estimated a residual contamination of 13$\%$ for the BSSs within one $r_{\rm h}$. As expected, the contamination of the BSSs alone is higher, since the MS of the SMC occupies the same region in the CMD.

The same Monte Carlo technique was adopted to estimate the residual contamination for NGC\,419, but with the additional complication given by the fact that this cluster is located along the Magellanic Bridge \citep{zivick2019}, and is thus contaminated by an additional population of stars that show a different kinematic behaviour from that of the SMC \citep{massari2021}. To take this complexity into account, we fitted the distributions of the two contaminant populations with distinct Gaussian functions. We then implemented the procedure described above separately for the two distributions. Adding up the contributions of the two populations, we obtained a value for the global residual contamination of 2$\%$ within one half-mass radius ($r_{\rm h}=36\farcs73 $, derived from $r_{\rm hl}$, see appendix \ref{app:RH}) in NGC 419. When limiting the analysis to the BSS region of the CMD, and thus selecting the sample of non-members as the sources with $m_{\mathrm{F555W}}<22.3$, $m_{\mathrm{F438W}}-m_{\mathrm{F814W}}<0.4$ and $r>r_{\rm c}=12\arcsec$ (\citealt{glatt2009}) the estimated contamination is $\sim4\%$. This value is significantly smaller than that obtained for NGC\,339, in agreement with our expectations, given the different underlying field population within the SMC field.

\section{Dynamical state of the clusters}\label{sec:dyn}
To study the dynamical state of these extra-Galactic clusters, we analyzed the level of segregation of their BSSs. We did so by calculating the $A^{+}_{rh}$ parameter, first defined by \cite{alessandrini2016}. In particular, following \cite{Lanzoni2016}, we determined the area enclosed between the cumulative radial distributions of BSSs, $\phi_{BSS}(x)$, and that of a reference (lighter) population, $\phi_{REF}(x)$, both measured within one half-mass radius:
\begin{equation}
    A^{+}_{rh}(x)=\int_{x_{\rm min}}^x \left (\, \phi_{BSS}(x') -\phi_{REF}(x') \, \right ) \, dx'
\end{equation}

\noindent where $x=log(r/r_{\rm h})$ is the cluster-centric distance expressed in logarithmic units, and $x_{\rm min}$ is the minimum value sampled (usually this coincides with the cluster centre, in which case its value is zero). The value of $A^{+}_{rh}$ increases with the level of segregation of the BSSs, effectively tracing the dynamical evolution of the cluster. This has been verified with numerical simulations \citep{alessandrini2016} and it has been proven to be effective for a conspicuous sample of old star clusters, both in the Milky Way and in the LMC (see \citealt{ferraro2018}; \citealt{FERRARO2019}). The value of this parameter is computed within $r_{\rm h}$ to allow a meaningful comparison among clusters with different structures and sizes.

\begin{figure}[h!]
\centering
\includegraphics[scale=0.3]{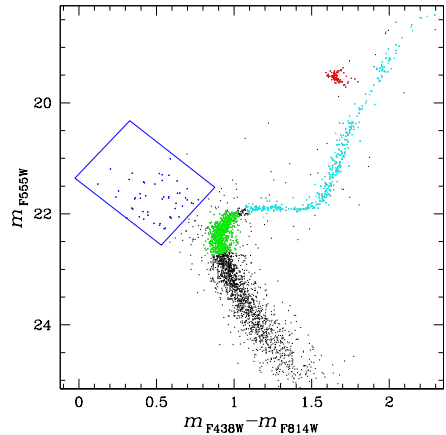}
\caption{Decontaminated CMD of NGC\,339. In blue we selected BSSs, in red, green and cyan the stars belonging to the RC, TO and RGB regions respectively.}
\label{fig:selezA339}
\end{figure}

Practically, to compute the value of $A^{+}_{rh}$ we first had to select from the CMD the member BSSs to analyze, along with the reference populations. As shown in Figure \ref{fig:selezA339}, we selected three different reference groups, TO stars in green, red giant branch (RGB) stars in cyan and red clump (RC) stars in red, and we estimated the value of $A^{+}_{rh}$ for the three cases. 
We remark at this point that our selection of BSSs is designed to achieve the best purity of the sample, at the cost of completeness. As an example, the BSSs selection was performed in different combinations of filters, and we only retained the stars located in the BSS region in all of these combinations. Moreover, the selection for NGC~419 likely misses the BSSs of the red part of the eMSTO, which are hidden among blue eMSTO stars. Yet, an attempt at recovering them would certainly drain-in contaminating non-BSSs, which we want to avoid.

\begin{figure}[h!]
\centering
\includegraphics[scale=0.29]{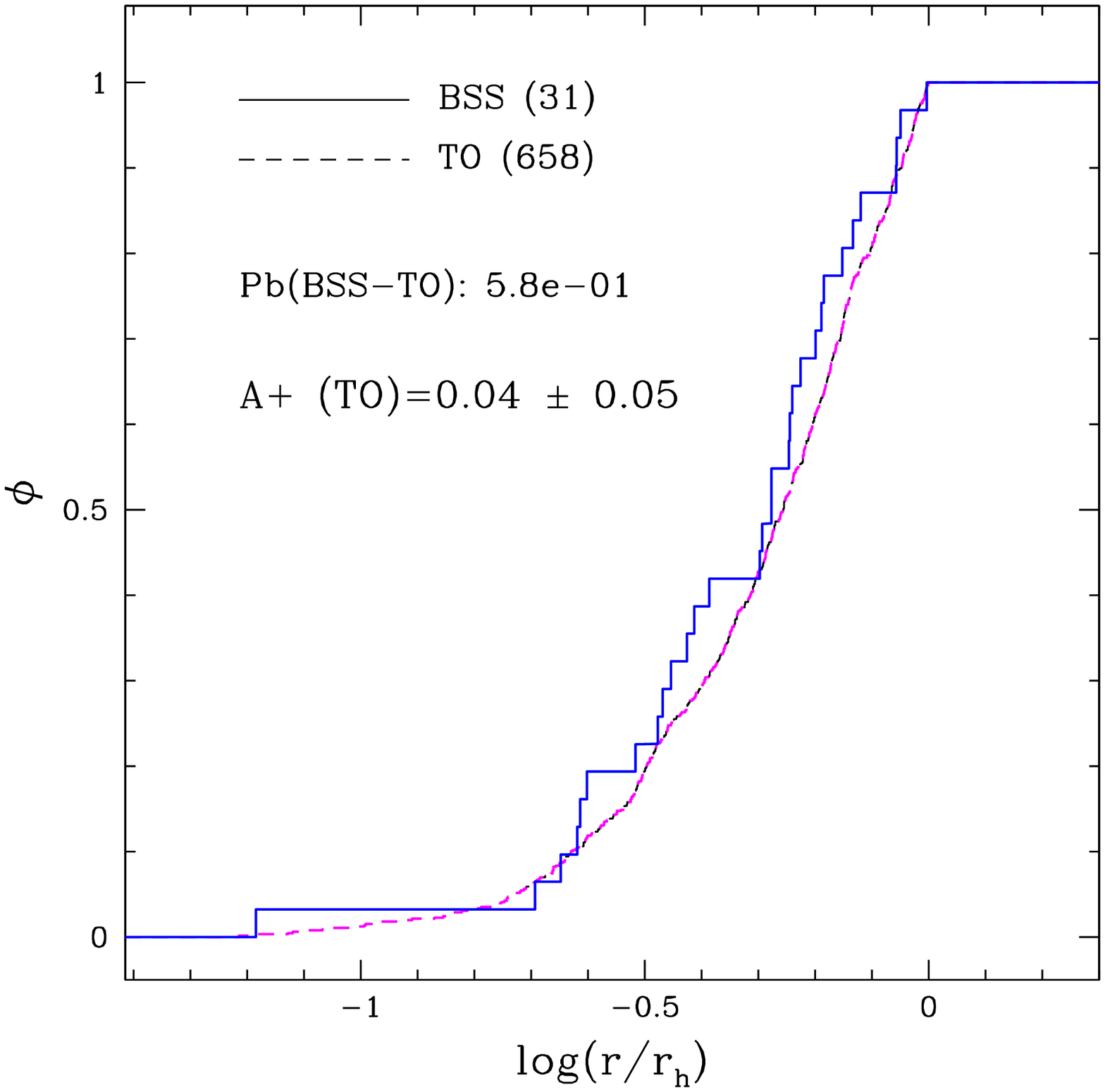}
\includegraphics[scale=0.29]{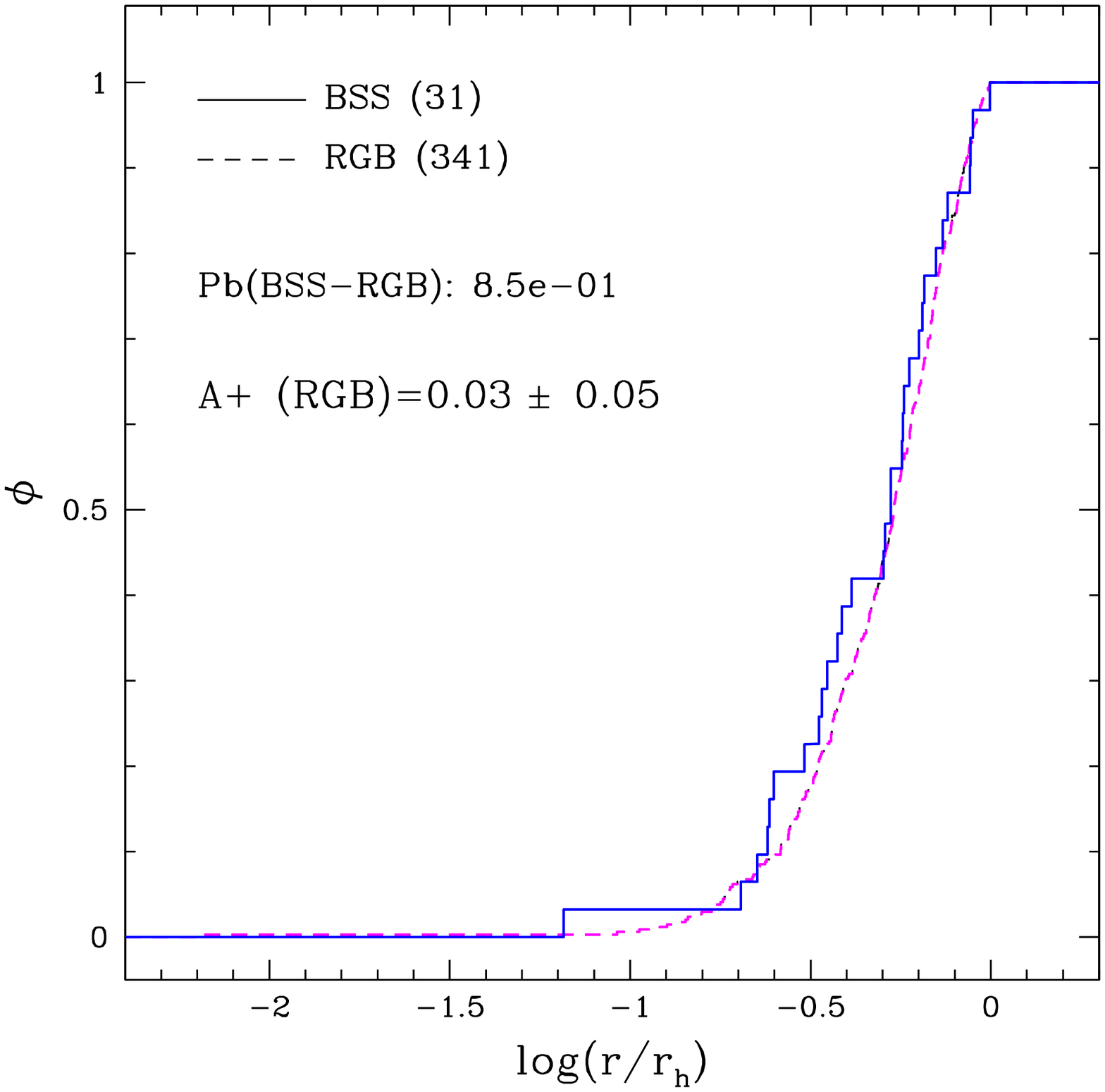}
\includegraphics[scale=0.29]{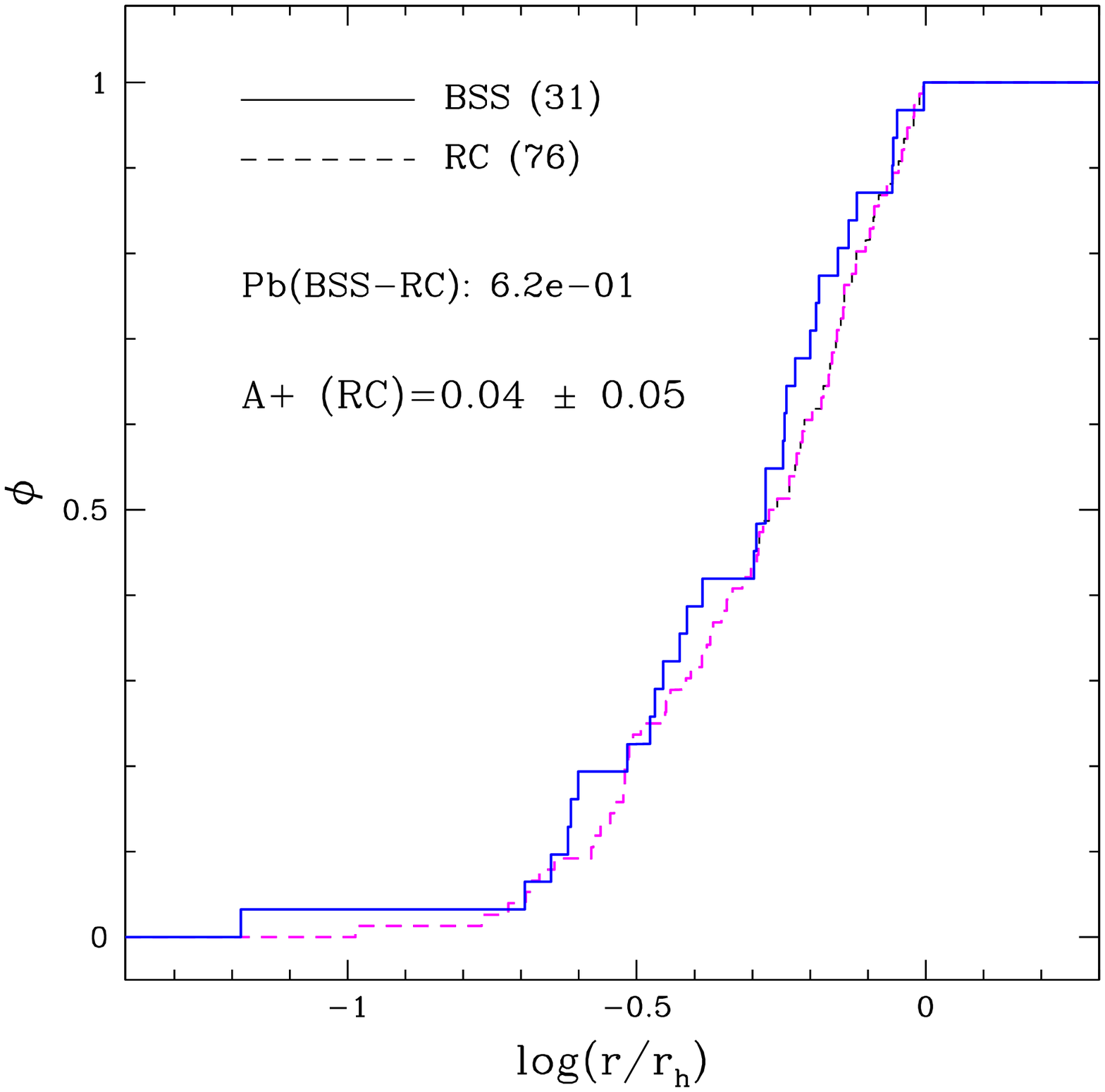}
\caption{Cumulative radial distributions of BSSs (blue lines) and of the reference stars (magenta lines) in NGC\,339. The number of stars in each population is labelled in all panels, together with the corresponding value of $A^{+}_{rh}$ and the KS probability that the two samples are drawn from the same parent family.}
\label{fig:distribA339}
\end{figure}

We built the cumulative radial distribution of BSSs and compared it with each of the reference populations as shown in Figure \ref{fig:distribA339}. The value of $A^{+}_{rh}$ has been determined for all of the three cases, together with its associated uncertainty. This has been estimated by using a jackknife bootstrapping technique \citep{Lupton93}. We also performed a Kolmogorov-Smirnov test to determine the level of confidence by which we can consider the two populations as different. 

In all cases (see Table \ref{tab:apiu339}) we find a value of $A^{+}_{rh}$ which is well consistent with zero within 1$\sigma$, indicating that the BSSs in NGC\,339 have yet to start their segregation, as expected for a dynamically young cluster, based on the evidence gathered from Galactic GCs. Consistently with the measured value of $A^{+}_{rh}$, the results of the KS-test show that the spatial distribution of the BSSs is statistically identical to that of the reference populations.

\begin{table}[h]
\centering
\setlength{\tabcolsep}{14pt}
\renewcommand{\arraystretch}{1.4}
\begin{center}
\begin{tabular}{ccccl}
\hline\hline
ref &$A^{+}_{rh}$&$\varepsilon_{A^{+}}$&$N_{\rm BSS}$& $N_{\rm ref}$\\ \hline 
TO&0.04&0.05&31&658\\ 
RGB&0.03&0.05&31&341\\ 
RC&0.04&0.05&31&76\\ \hline
\end{tabular}
\caption{Values of $A^{+}_{rh}$ (column 2) and their error $\varepsilon_{A^{+}}$ (column 3), computed for the different reference populations (column 1). $N_{\rm BSS}$ and $N_{\rm ref}$ indicate the number of stars considered by the selection of BSSs and of the reference population in NGC\,339, respectively.}
\label{tab:apiu339}
\end{center}
\renewcommand{\arraystretch}{1.0}
\end{table}

\begin{figure}[h!]
\centering
\includegraphics[scale=0.3]{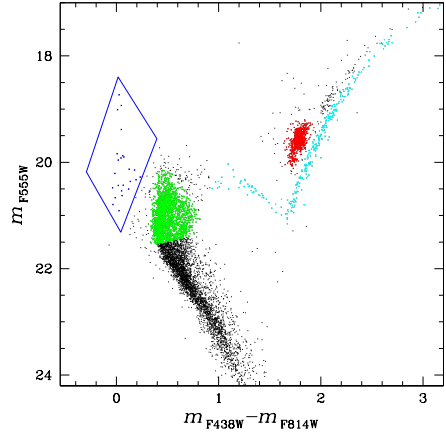}
\caption{Decontaminated CMD of NGC\,419. In blue we selected BSSs, in red, green and cyan the stars belonging to the RC, TO and RGB regions respectively.}
\label{fig:selezA419}
\end{figure}

\begin{figure}[h!]
\centering
\includegraphics[scale=0.29]{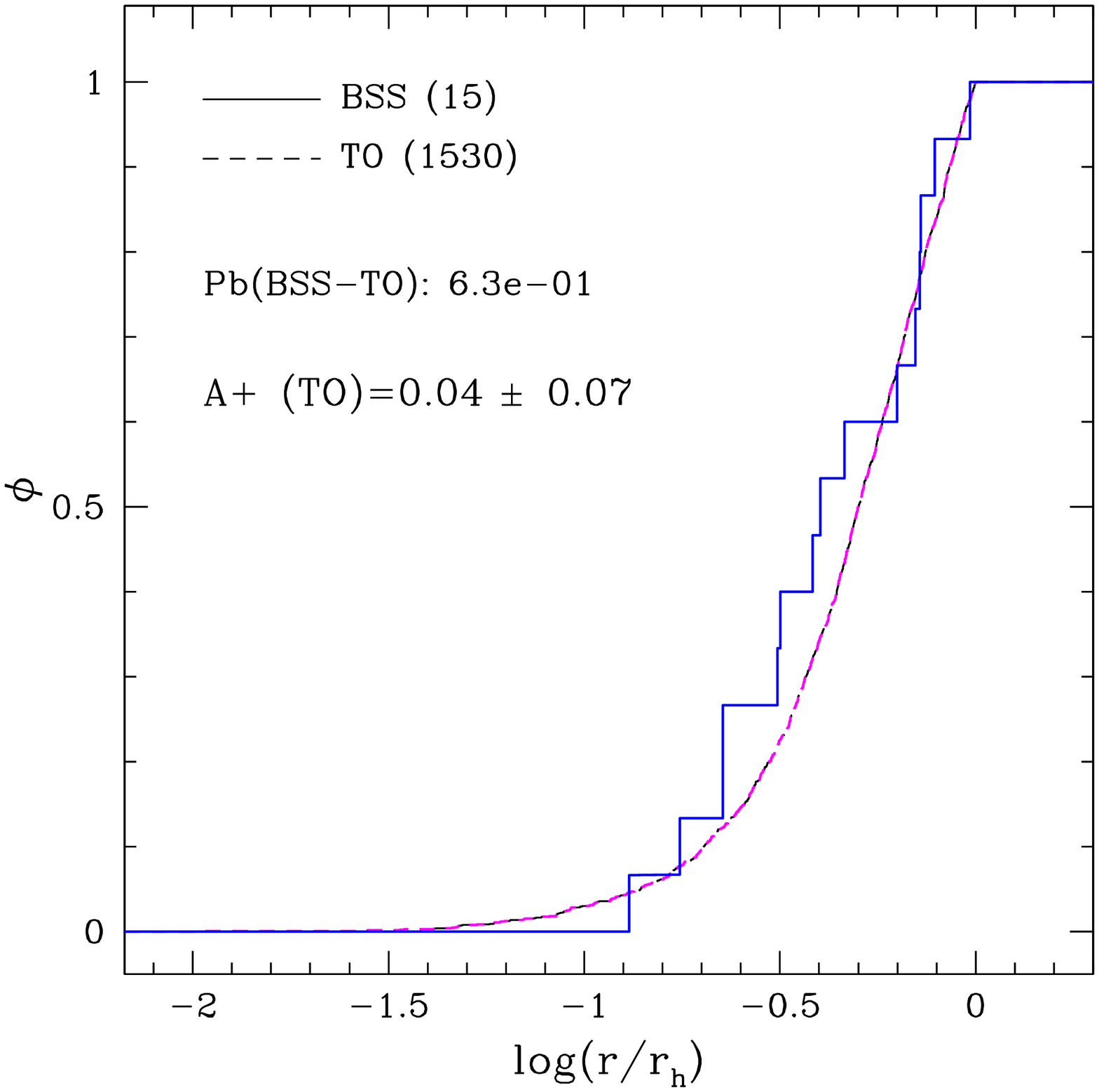}
\includegraphics[scale=0.29]{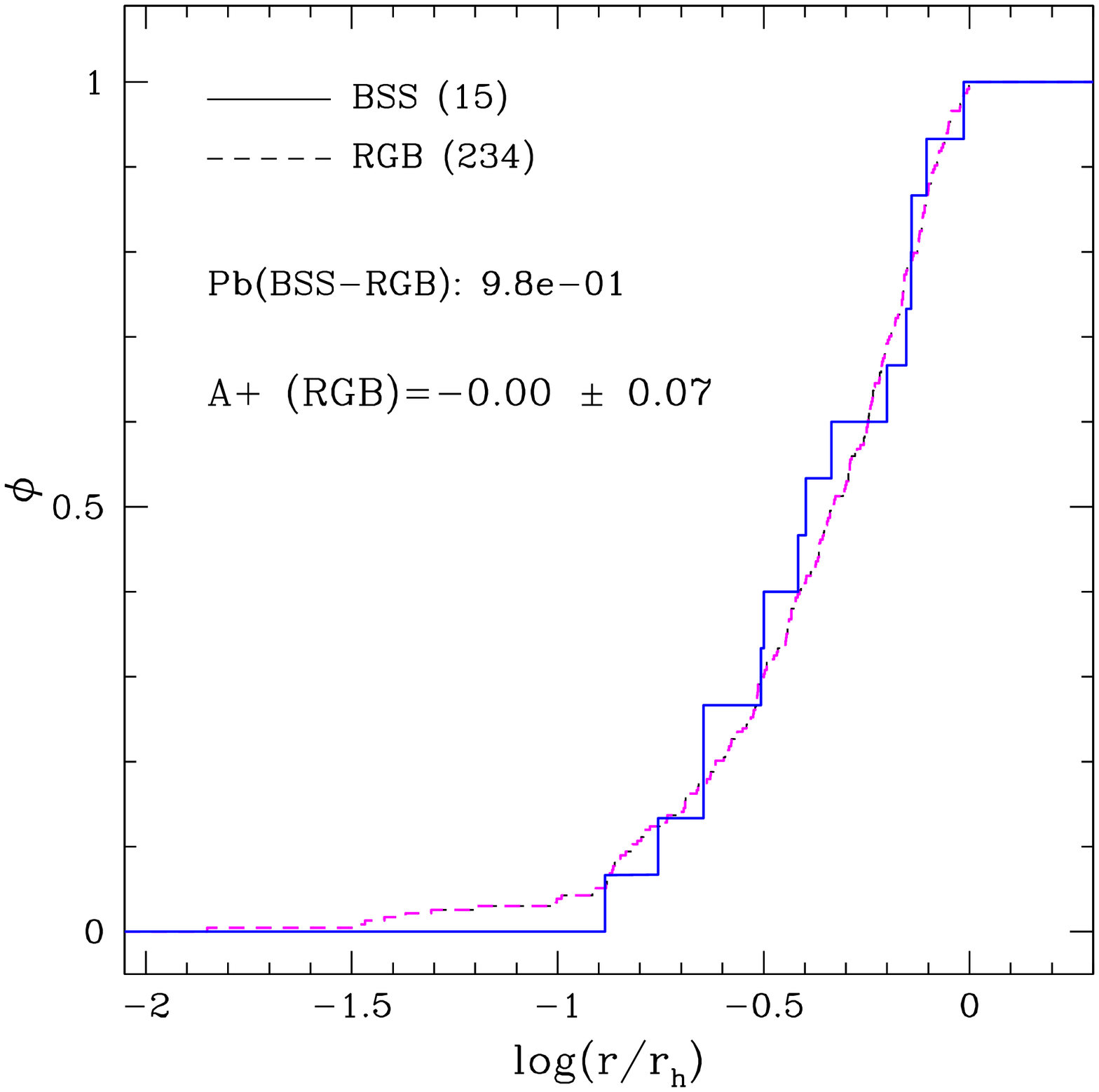}
\includegraphics[scale=0.29]{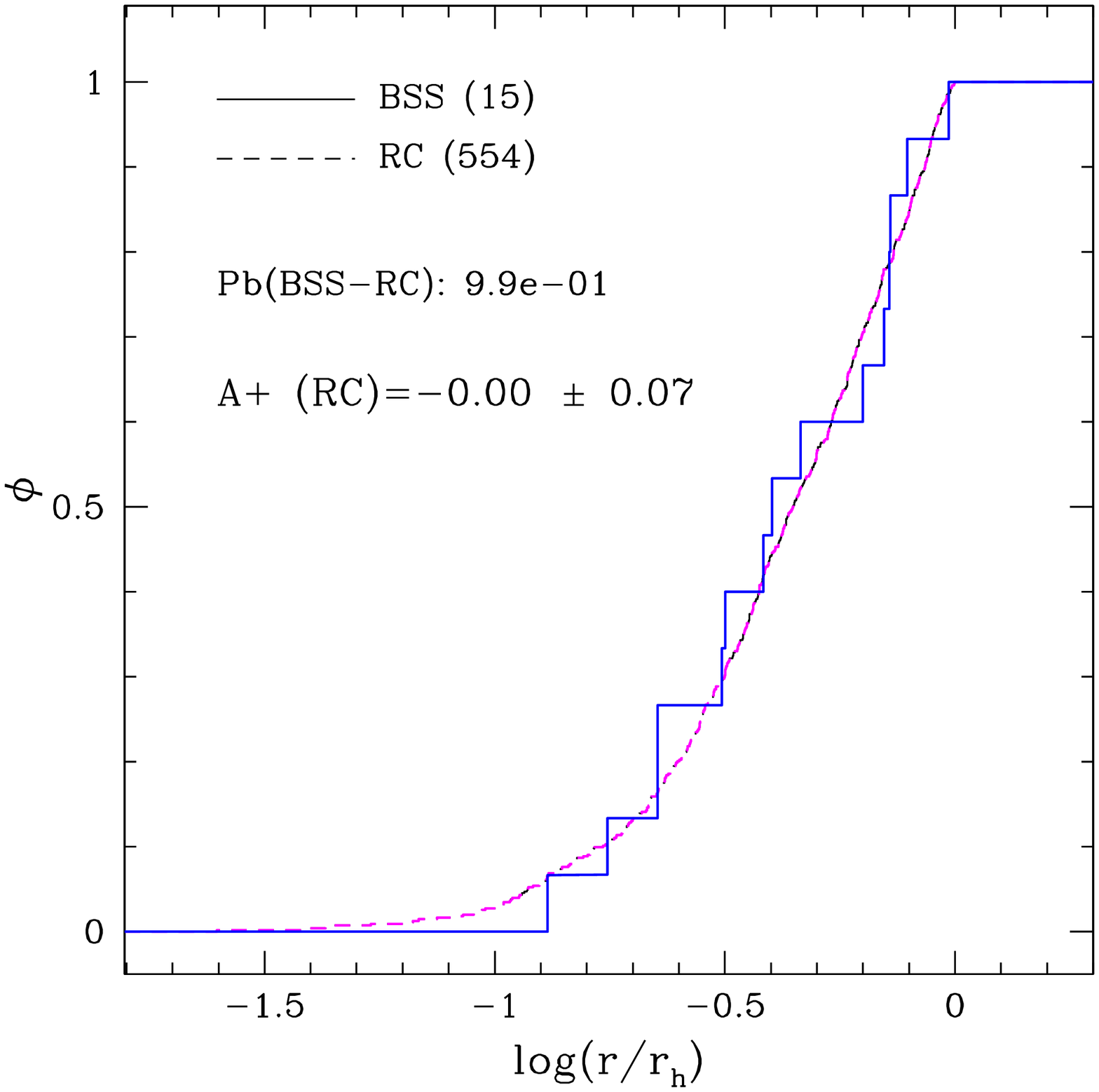}
\caption{Similar to Figure \ref{fig:distribA339}, but for NGC\,419.}
\label{fig:distribA419}
\end{figure}

The study of the dynamical state of NGC\,419 has been performed in the same way, starting from the CMD selection of the stellar populations (Figure \ref{fig:selezA419}).  The cumulative radial distributions are analyzed out to $r_{\rm h}=36\farcs73$ and are plotted in Figure \ref{fig:distribA419}. The values of $A^{+}_{rh}$ and the errors computed for this cluster (Table \ref{tab:apiu419}) are also consistent with zero within $1\sigma$, meaning that, even in the case of NGC\,419, the BSS population is not segregated. This is again confirmed by the results of the KS-test.

\begin{table}[h!]
\centering
\setlength{\tabcolsep}{14pt}
\renewcommand{\arraystretch}{1.4}
\begin{center}
\begin{tabular}{ccccl}
\hline\hline
ref&$A^{+}_{rh}$&$\varepsilon_{A^{+}}$&$N_{\rm BSS}$& $N_{\rm ref}$\\ \hline 
TO&0.04&0.07&15&1530\\ 
RGB&0.00&0.07&15&234\\ 
RC&0.00&0.07&15&554\\ \hline
\end{tabular}
\caption{Values of $A^{+}_{rh}$ (column 2) and their error $\varepsilon_{A^{+}}$ (column 3), computed for the different reference populations (column 1). $N_{\rm BSS}$ and $N_{\rm ref}$ indicate the number of stars considered by the selection of BSSs and of the reference population in NGC\,419, respectively.}
\label{tab:apiu419}
\end{center}
\renewcommand{\arraystretch}{1.0}
\end{table}

For both clusters, the value of $A^{+}_{rh}$ suggests that their BSS populations are not yet centrally segregated. To validate that conclusion is robust, we further performed several checks. First, we tested the effect of residual field contamination on the values of $A^{+}_{rh}$. For NGC\,339, for example, the estimated contamination of 13$\%$ within $r_{\rm h}$ implies that 4 out of 31 the identified BSSs are statistically not members. We therefore computed a new value of $A^{+}_{rh}$ after randomly removing 4 stars, spatially distributed as field sources (with a uniform spatial distribution), from the BSS distribution.
By recomputing $A^{+}_{rh}$ for this new distributions of 27 BSS and repeating the operation 30 times, we obtained an average value of $A^{+}_{rh}\sim 0.05$ (with rms $=0.01$), which is perfectly consistent with our original results and thus demonstrates that the effect of the residual contamination is negligible.
We also studied whether the quality selection procedure could influence our final results. To do so we estimated the values of $A^{+}_{rh}$ on a catalogue of NGC\,339 stars that were not selected based on the quality criteria described in Section \ref{subsec:criteria} but only with the $2\sigma$ kinematic selection. We obtained the new value $A^{+}_{rh}\sim 0.03 \pm 0.04$ that is consistent with the previous values, proving that the technique used to remove low quality sources didn't significantly alter the final result. On the other hand, we noticed a significant difference when the kinematic selection of stellar members is not considered. In fact, we obtained a value of $A^{+}_{rh}\sim -0.06 \pm 0.02$, which is not consistent with our previous results and can only be explained as due to the contamination of the BSS sample by a less concentrate population, like SMC stars. This result also proves that the kinematic selection is essential to correctly remove the contamination from the Cloud.

To investigate the connection between the BSS segregation level and the dynamical status of each cluster, we studied the relation between the measured values of $A^{+}_{rh}$ and the dynamical/structural properties of the systems. For $\sim$1/3 of the entire population of Galactic GCs, \cite{ferraro2018} found a correlation between the value of $A^{+}_{rh}$ and the number of current central relaxation times experienced by the systems since formation, as expressed by the parameter $N _{\rm relax} = t /t_{\rm rc}$, where $t$ is the cluster age and $t_{\rm rc}$ is its central relaxation time. The value of $A^{+}_{rh}$ also shows an anti-correlation with the cluster core radius, in agreement with the fact that dynamically more evolved clusters have smaller $r_{\rm c}$ and higher levels of segregation. The same relations were also found in old extra-Galactic clusters, situated in the Large Magellanic Cloud \citep{FERRARO2019}. 

All the stellar systems that have been studied until now using these techniques, Galactic and extra-Galactic, are chronologically old, while the systems studied in this paper are significantly younger. To verify whether the same relations hold also for these young systems, we determined their central relaxation time $t_{\rm rc}$, following \cite{djo}:
\begin{equation}
    t_{\rm rc} =1.491 \cdot 10^{7}\, \mathrm{yr} \,\frac{0.5592}{\ln (0.4\,\it{N})}\left<m\right>^{-1} \rho_{M,0}^{1/2} \ r_{\rm c}^3
    \label{eqdjo2}
\end{equation}
where $N$ is the estimated total number of stars, $\left \langle m\right \rangle$ is the average stellar mass in solar units, $\rho_{M,0}$ is the central mass density in $\rm{M_{\odot}/pc^3}$ and $r_{\rm c}$ is the core radius in pc. For NGC\,339, $\log(\rho_{M,0})= 1.29$ and $r_{\rm c}=7.38$ pc \citep{MACKEY}, while the average stellar mass ($\left \langle m\right \rangle=0.3M_{\odot}$) was computed from a synthetic population created by using the evolutionary models presented in \cite{marigo2008}\footnote{\url{http://stev.oapd.inaf.it/cgi-bin/cmd}} and the following physical properties: age t=6\,Gyr and metallicity Z=0.001 (\citealt{glatt2009}). The total number of stars ($N$) has then been derived as the ratio between the total cluster mass ($M\sim 10^5 M_{\odot}$; \citealt{MACKEY}) and $\left \langle m\right \rangle$. Once all the parameters were known, we determined the central relaxation time $t_{\rm rc}=4.16$ Gyr, implying $N_{\rm relax}=t /t_{\rm rc}=1.44$. Such a small value of $N_{\rm relax}$ is perfectly consistent with the conclusion derived from the $A^{+}_{rh}$ parameter that this cluster is dynamically young. For NGC\,419, we determined the average stellar mass in a similar way, obtaining $\left \langle m\right \rangle=0.358\rm{M_{\odot}}$ (from t=1.5 Gyr, Z=0.004, \citealt{glatt2009}; $\rm{M} \sim 0.8 \cdot 10^5 \ \rm{M_{\odot}}$, \citealt{song2019}). Not all the parameters needed for the calculation are known and so, to determine the central mass density  ($\rho_{M,0}$), we started by estimating the values of the color excess and the distance module. 
\begin{figure}[h!]
\centering
\includegraphics[scale=0.3]{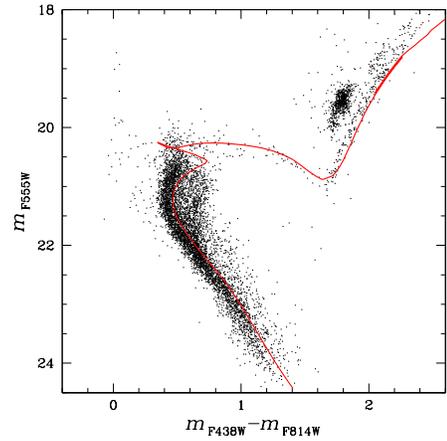}
\caption{CMD of NGC\,419 and the isochrone computed for t=1.5\,Gyr and Z=0.004 \citep{marigo2008}, superposed as a red line.}
\label{fig:isocrona}
\end{figure}
We fitted by eye the isochrone curve, built with the previous evolutionary models, on the CMD of the cluster (Figure \ref{fig:isocrona}), so that it best described the behaviour of all the evolutionary sequences. In this way, we were able to estimate the following values: $E(B-V)=0.08$ e $(m-M)_0 =18.94$. By following the prescriptions given in \cite{djo} we computed, with these quantities, the surface brightness in units of luminosity per parsec square ($ \mathrm{L/pc^2}$, eq. 5) and determined the $p$ parameter (eq. 6). With these values, we determined the central luminosity density (eq. 4), from which we obtained the central mass density $\rho_{M.0}$ using the $M/L$ ratio. Finally, we determined the central relaxation time: $t_{\rm rc}=1.56$ Gyr, which leads to $N_{\rm relax}=t /t_{\rm rc}=0.962$.

This value of $N_{\rm relax}$ confirms that the system is dynamically young and this theoretical prediction based on the structural properties of the cluster agrees with the results obtained from the observational study of the BSSs.

\begin{table}[h]
\setlength{\tabcolsep}{3pt}
\renewcommand{\arraystretch}{1.25}
\begin{center}
\begin{tabular}{ c c c c c c}
\hline\hline
$t$& c & $r_{\rm c}$ & $\mu_{555}(0)$ & $M_V$ & M/L\\ 
(Gyr) && (arcsec) & (mag $\mathrm{arcsec^{-2}}$)& (mag) & ($M_{\odot} L_{\odot}^{-1}$) \\ \hline
\renewcommand{\arraystretch}{1.4}
1.5& 1.059 &  15.22 & 18.18 & $-8.85$ & 0.22\\ 
(1)&(1)& (1)&(1)&(1)&(2)\\
\hline
\end{tabular}
\caption{Parameters of NGC\,419. References: \cite{glatt2009} (1) and \cite{song2019} (2). }
\label{tab:dataDJ}
\end{center}
\renewcommand{\arraystretch}{1.0}
\end{table}

\begin{figure}[h!]
\centering
\includegraphics[scale=0.38]{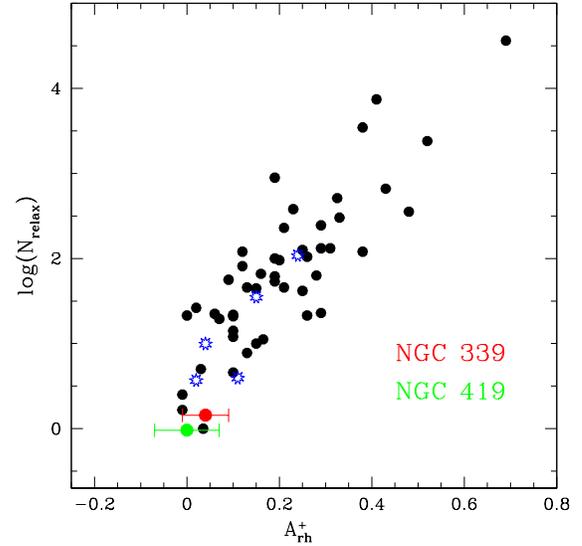}
\caption{Relation between the number of relaxations $N_{\rm relax}=t/t_{\rm rc}$ and the $A^{+}_{rh}$ parameter for Galactic clusters (black), for LMC clusters (blue) and for the two systems analyzed in this work.}
\label{fig:Nrelfinali}
\end{figure}

\begin{figure}[h!]
\centering
\includegraphics[scale=0.38]{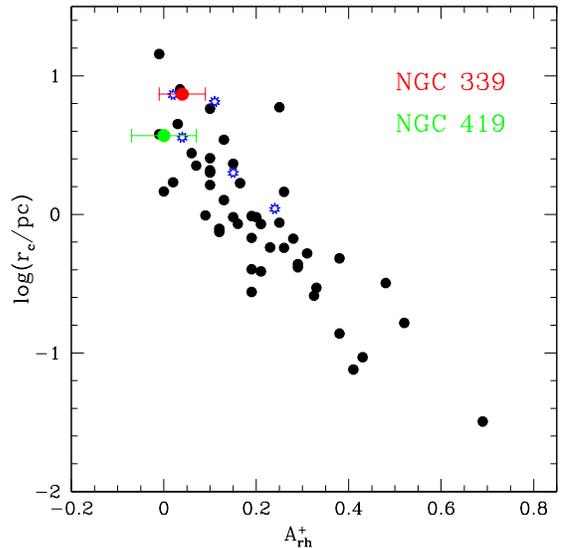}
\caption{Relation between the core radius $r_{\rm c}$ and the $A^{+}_{rh}$ parameter for star clusters in the MW (black), in the LMC (blue) and the two systems analyzed in this work (red and green circles; see labels).}
\label{fig:rcfinali}
\end{figure}

Figures \ref{fig:Nrelfinali} and \ref{fig:rcfinali} summarise the main results of our work. In the first plot we present the relation between $A^{+}_{rh}$ and the number of relaxations ${N_{\rm relax}}$ obtained from the study of star clusters in the Milky Way (\citealt{ferraro2018}, black symbols) and in the Large Magellanic Cloud (\citealt{FERRARO2019}, blue symbols). The two SMC clusters analyzed here are positioned along the same sequence in which dynamically young clusters are expected to be located. This means that the study of the dynamical state of clusters based on the radial distribution of BSSs is effective also in this extra-Galactic environment and, especially, it is valid even for clusters that are young or of intermediate age.

The same result is found by studying the trend of $A^{+}_{rh}$ as a function of the core radii of the clusters. From the plot in Figure \ref{fig:rcfinali}, NGC\,339 and NGC\,419 are located precisely along the sequence of Galactic and LMC clusters, despite being younger and in an extra-Galactic system.

\section{Conclusions}\label{sec:conclusio}
In this paper we analyzed multi-epoch \textit{HST} data of two stellar clusters, NGC\,339 and NGC\,419 with the aim of determining the stage of the dynamical evolution of these two young systems by using the “dynamical clock" (\citealt{Ferraro2012}, \citealt{Lanzoni2016}). In doing this we followed the approach proposed by \cite{Lanzoni2016} and \cite{ferraro2018} based on the  measure of  the level of segregation of the BSS population via the $A^{+}_{rh}$ parameter (see also \citealt{alessandrini2016}). After a kinematic selection of Blue Straggler stars and an estimation of their residual contamination, we built cumulative radial distributions of BSSs and of a few reference populations and we found values of $A^{+}_{rh}$ consistent with zero for both clusters, indicative of an absence of segregation. By comparing these results with the dynamical properties of the systems (central relaxation times and core radii), we determined that both clusters are dynamically young and that they show the same correlations found in old star clusters from the MW and the LMC, confirming that the $A^{+}_{rh}$ parameter is an efficient hand of the “dynamical clock" even for young and intermediate age clusters.

\

We thank the anonymous referee for comments and suggestions that improved the quality of our paper. This research is part of the project Cosmic-Lab (“Globular Clusters as
Cosmic Laboratories”) at the Physics and Astronomy Department of the Bologna University (see the web page: \url{http://www.cosmic-lab.eu/Cosmic-Lab/Home.html}). The research is funded by the project Light-on-Dark granted by MIUR through PRIN2017K7REXT contract (PI: Ferraro). This work is also based on observations made with the NASA/ESA Hubble Space Telescope, obtained from the Data Archive at the Space Telescope Science Institute, which is operated by the Association of Universities for Research in Astronomy, Inc., under NASAcontract NAS 5-26555.

\newpage

\appendix
\section{The half-mass radius}\label{app:RH}
\begin{figure}[h!]
\centering
\includegraphics[scale=0.31]{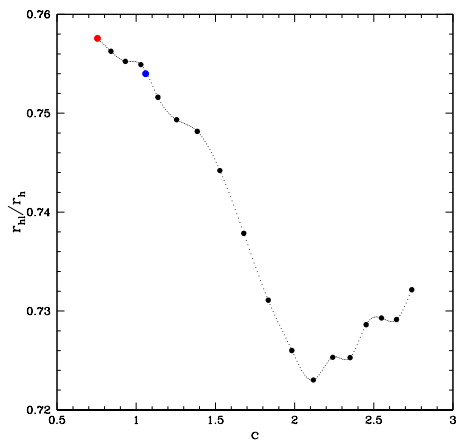}
\caption{Trend of $r_{\rm hl}/r_{\rm h}$ with the value of concentration $c$. The red dot represents the values NGC\,339, the blue dot the values for NGC\,419.}
\label{fig:raggio3d}
\end{figure}

The half-mass radius is a three-dimensional parameter and as such it cannot be directly measured from observations. To determine the values of $r_{\rm h}$ for NGC\,339, we used the projected half-light radius $\rm{r_{\rm hl}}=52\farcs10$, measured by \cite{glatt2011}, and we took advantage of the relation existing between this two value when a King model is assumed to be a good representation of the observed density profile. Using a dedicated online software\footnote{\url{http://www.cosmic-lab.eu/bhking/index.php}} \citep{miocchi2013} we built several King models with different values of the concentration parameter $c$. For each one of these models the values of $r_{\rm h}$ and $r_{\rm hl}$ are known, so we can obtain a relation between $c$ and the ratio $r_{\rm hl}/r_{\rm h}$. The trend of this relation is shown in Figure \ref{fig:raggio3d}. For NGC\,339, c=0.755 \citep{glatt2011}, so that we had to extrapolate the value $r_{\rm hl}/r_{\rm h}=0.757$, that corresponds to a half-mass radius of $r_{\rm h}=68\farcs78$.

\noindent For NGC\,419, \cite{glatt2009} determined  $r_{\rm hl}= 27\farcs69$ and c=1.059, so we obtained  $r_{\rm h}=36\farcs73$.

\newpage

\bibliographystyle{aasjournal}
\bibliography{Bibliography}

\begin{thebibliography}{}
\expandafter\ifx\csname natexlab\endcsname\relax\def\natexlab#1{#1}\fi
\providecommand{\url}[1]{\href{#1}{#1}}
\providecommand{\dodoi}[1]{doi:~\href{http://doi.org/#1}{\nolinkurl{#1}}}
\providecommand{\doeprint}[1]{\href{http://ascl.net/#1}{\nolinkurl{http://ascl.net/#1}}}
\providecommand{\doarXiv}[1]{\href{https://arxiv.org/abs/#1}{\nolinkurl{https://arxiv.org/abs/#1}}}

\bibitem[{{Alessandrini} {et~al.}(2016){Alessandrini}, {Lanzoni}, {Ferraro},
  {Miocchi}, \& {Vesperini}}]{alessandrini2016}
{Alessandrini}, E., {Lanzoni}, B., {Ferraro}, F.~R., {Miocchi}, P., \&
  {Vesperini}, E. 2016, \apj, 833, 252, \dodoi{10.3847/1538-4357/833/2/252}

\bibitem[{{Anderson} \& {Bedin}(2010)}]{2010andersonBedin}
{Anderson}, J., \& {Bedin}, L.~R. 2010, \pasp, 122, 1035,
  \dodoi{10.1086/656399}

\bibitem[{{Anderson} \& {King}(2006)}]{andersonking2006}
{Anderson}, J., \& {King}, I.~R. 2006, {PSFs, Photometry, and Astronomy for the
  ACS/WFC}, Instrument Science Report ACS 2006-01

\bibitem[{{Bedin} {et~al.}(2008){Bedin}, {King}, {Anderson}, {Piotto},
  {Salaris}, {Cassisi}, \& {Serenelli}}]{bedin}
{Bedin}, L.~R., {King}, I.~R., {Anderson}, J., {et~al.} 2008, \apj, 678, 1279,
  \dodoi{10.1086/529370}

\bibitem[{{Bellini} {et~al.}(2011){Bellini}, {Anderson}, \&
  {Bedin}}]{belliniAndersonBedin2011}
{Bellini}, A., {Anderson}, J., \& {Bedin}, L.~R. 2011, \pasp, 123, 622,
  \dodoi{10.1086/659878}

\bibitem[{{Bellini} {et~al.}(2017{\natexlab{a}}){Bellini}, {Anderson}, {Bedin},
  {King}, {van der Marel}, {Piotto}, \& {Cool}}]{Bellini2017a}
{Bellini}, A., {Anderson}, J., {Bedin}, L.~R., {et~al.} 2017{\natexlab{a}},
  \apj, 842, 6, \dodoi{10.3847/1538-4357/aa7059}

\bibitem[{{Bellini} {et~al.}(2017{\natexlab{b}}){Bellini}, {Bianchini},
  {Varri}, {Anderson}, {Piotto}, {van der Marel}, {Vesperini}, \&
  {Watkins}}]{bellini2017}
{Bellini}, A., {Bianchini}, P., {Varri}, A.~L., {et~al.} 2017{\natexlab{b}},
  \apj, 844, 167, \dodoi{10.3847/1538-4357/aa7c5f}

\bibitem[{{Bellini} {et~al.}(2014){Bellini}, {Anderson}, {van der Marel},
  {Watkins}, {King}, {Bianchini}, {Chanam{\'e}}, {Chandar}, {Cool}, {Ferraro},
  {Ford}, \& {Massari}}]{Bellini20142014}
{Bellini}, A., {Anderson}, J., {van der Marel}, R.~P., {et~al.} 2014, \apj,
  797, 115, \dodoi{10.1088/0004-637X/797/2/115}

\bibitem[{{Bellini} {et~al.}(2018){Bellini}, {Libralato}, {Bedin}, {Milone},
  {van der Marel}, {Anderson}, {Apai}, {Burgasser}, {Marino}, \&
  {Rees}}]{bellini2018}
{Bellini}, A., {Libralato}, M., {Bedin}, L.~R., {et~al.} 2018, \apj, 853, 86,
  \dodoi{10.3847/1538-4357/aaa3ec}

\bibitem[{{Cabrera-Ziri} {et~al.}(2016){Cabrera-Ziri}, {Niederhofer},
  {Bastian}, {Rejkuba}, {Balbinot}, {Kerzendorf}, {Larsen}, {Mackey},
  {Dalessandro}, {Mucciarelli}, {Charbonnel}, {Hilker}, {Gieles}, \&
  {H{\'e}nault-Brunet}}]{CABRERA}
{Cabrera-Ziri}, I., {Niederhofer}, F., {Bastian}, N., {et~al.} 2016, \mnras,
  459, 4218, \dodoi{10.1093/mnras/stw966}

\bibitem[{{Cioni} {et~al.}(2000){Cioni}, {van der Marel}, {Loup}, \&
  {Habing}}]{cioni2000}
{Cioni}, M. R.~L., {van der Marel}, R.~P., {Loup}, C., \& {Habing}, H.~J. 2000,
  \aa, 359, 601.
\newblock \doarXiv{astro-ph/0003223}

\bibitem[{{Dalessandro} {et~al.}(2019){Dalessandro}, {Ferraro}, {Bastian},
  {Cadelano}, {Lanzoni}, \& {Raso}}]{dalessandro2019}
{Dalessandro}, E., {Ferraro}, F.~R., {Bastian}, N., {et~al.} 2019, \aap, 621,
  A45, \dodoi{10.1051/0004-6361/201834011}

\bibitem[{{Djorgovski}(1993)}]{djo}
{Djorgovski}, S. 1993, in Astronomical Society of the Pacific Conference
  Series, Vol.~50, Structure and Dynamics of Globular Clusters, ed. S.~G.
  {Djorgovski} \& G.~{Meylan}, 373

\bibitem[{Ferraro {et~al.}(1992)Ferraro, Fusi~Pecci, \& Buonanno}]{Ferraro1992}
Ferraro, F.~R., Fusi~Pecci, F., \& Buonanno, R. 1992, Monthly Notices of the
  Royal Astronomical Society, 256, 376, \dodoi{10.1093/mnras/256.3.376}

\bibitem[{{Ferraro} {et~al.}(2019){Ferraro}, {Lanzoni}, {Dalessandro},
  {Cadelano}, {Raso}, {Mucciarelli}, {Beccari}, \& {Pallanca}}]{FERRARO2019}
{Ferraro}, F.~R., {Lanzoni}, B., {Dalessandro}, E., {et~al.} 2019, Nature
  Astronomy, 3, 1149, \dodoi{10.1038/s41550-019-0865-1}

\bibitem[{{Ferraro} {et~al.}(1993){Ferraro}, {Pecci}, {Cacciari}, {Corsi},
  {Buonanno}, {Fahlman}, \& {Richer}}]{Ferraro1993}
{Ferraro}, F.~R., {Pecci}, F.~F., {Cacciari}, C., {et~al.} 1993, \aj, 106,
  2324, \dodoi{10.1086/116804}

\bibitem[{{Ferraro} {et~al.}(1997){Ferraro}, {Paltrinieri}, {Fusi Pecci},
  {Cacciari}, {Dorman}, {Rood}, {Buonanno}, {Corsi}, {Burgarella}, \&
  {Laget}}]{ferraro1997}
{Ferraro}, F.~R., {Paltrinieri}, B., {Fusi Pecci}, F., {et~al.} 1997, \aap,
  324, 915.
\newblock \doarXiv{astro-ph/9703026}

\bibitem[{{Ferraro} {et~al.}(2012){Ferraro}, {Lanzoni}, {Dalessandro},
  {Beccari}, {Pasquato}, {Miocchi}, {Rood}, {Sigurdsson}, {Sills}, {Vesperini},
  {Mapelli}, {Contreras}, {Sanna}, \& {Mucciarelli}}]{Ferraro2012}
{Ferraro}, F.~R., {Lanzoni}, B., {Dalessandro}, E., {et~al.} 2012, \nat, 492,
  393, \dodoi{10.1038/nature11686}

\bibitem[{{Ferraro} {et~al.}(2018){Ferraro}, {Lanzoni}, {Raso}, {Nardiello},
  {Dalessandro}, {Vesperini}, {Piotto}, {Pallanca}, {Beccari}, {Bellini},
  {Libralato}, {Anderson}, {Aparicio}, {Bedin}, {Cassisi}, {Milone},
  {Ortolani}, {Renzini}, {Salaris}, \& {van der Marel}}]{ferraro2018}
{Ferraro}, F.~R., {Lanzoni}, B., {Raso}, S., {et~al.} 2018, \apj, 860, 36,
  \dodoi{10.3847/1538-4357/aac01c}

\bibitem[{{Fiorentino} {et~al.}(2014){Fiorentino}, {Lanzoni}, {Dalessandro},
  {Ferraro}, {Bono}, \& {Marconi}}]{2014fiore}
{Fiorentino}, G., {Lanzoni}, B., {Dalessandro}, E., {et~al.} 2014, \apj, 783,
  34, \dodoi{10.1088/0004-637X/783/1/34}

\bibitem[{{Glatt} {et~al.}(2009){Glatt}, {Grebel}, {Gallagher}, {Nota},
  {Sabbi}, {Sirianni}, {Clementini}, {Da Costa}, {Tosi}, {Harbeck}, {Koch}, \&
  {Kayser}}]{glatt2009}
{Glatt}, K., {Grebel}, E.~K., {Gallagher}, John~S., I., {et~al.} 2009, \aj,
  138, 1403, \dodoi{10.1088/0004-6256/138/5/1403}

\bibitem[{{Glatt} {et~al.}(2011){Glatt}, {Grebel}, {Jordi}, {Gallagher}, {Da
  Costa}, {Clementini}, {Tosi}, {Harbeck}, {Nota}, {Sabbi}, \&
  {Sirianni}}]{glatt2011}
{Glatt}, K., {Grebel}, E.~K., {Jordi}, K., {et~al.} 2011, \aj, 142, 36,
  \dodoi{10.1088/0004-6256/142/2/36}

\bibitem[{{Hills} \& {Day}(1976)}]{1976hills}
{Hills}, J.~G., \& {Day}, C.~A. 1976, \aplett, 17, 87

\bibitem[{{Lanzoni} {et~al.}(2016){Lanzoni}, {Ferraro}, {Alessandrini},
  {Dalessandro}, {Vesperini}, \& {Raso}}]{Lanzoni2016}
{Lanzoni}, B., {Ferraro}, F.~R., {Alessandrini}, E., {et~al.} 2016, \apj, 833,
  L29, \dodoi{10.3847/2041-8213/833/2/L29}

\bibitem[{{Libralato} {et~al.}(2019){Libralato}, {Bellini}, {Piotto},
  {Nardiello}, {van der Marel}, {Anderson}, {Bedin}, \&
  {Vesperini}}]{libralato2019}
{Libralato}, M., {Bellini}, A., {Piotto}, G., {et~al.} 2019, \apj, 873, 109,
  \dodoi{10.3847/1538-4357/ab0551}

\bibitem[{{Libralato} {et~al.}(2018){Libralato}, {Bellini}, {van der Marel},
  {Anderson}, {Watkins}, {Piotto}, {Ferraro}, {Nardiello}, \&
  {Vesperini}}]{libralato2018}
{Libralato}, M., {Bellini}, A., {van der Marel}, R.~P., {et~al.} 2018, \apj,
  861, 99, \dodoi{10.3847/1538-4357/aac6c0}

\bibitem[{{Lupton}(1993)}]{Lupton93}
{Lupton}, R. 1993, {Statistics in Theory and Practice} (Princeton, NJ:
  Princeton Univ. Press)

\bibitem[{{Mackey} \& {Gilmore}(2003)}]{MACKEY}
{Mackey}, A.~D., \& {Gilmore}, G.~F. 2003, \mnras, 338, 120,
  \dodoi{10.1046/j.1365-8711.2003.06022.x}

\bibitem[{{Marigo} {et~al.}(2008){Marigo}, {Girardi}, {Bressan}, {Groenewegen},
  {Silva}, \& {Granato}}]{marigo2008}
{Marigo}, P., {Girardi}, L., {Bressan}, A., {et~al.} 2008, \aa, 482, 883,
  \dodoi{10.1051/0004-6361:20078467}

\bibitem[{{Massari} {et~al.}(2021){Massari}, {Raso}, {Libralato}, \&
  {Bellini}}]{massari2021}
{Massari}, D., {Raso}, S., {Libralato}, M., \& {Bellini}, A. 2021, \mnras, 500,
  2012, \dodoi{10.1093/mnras/staa3497}

\bibitem[{{McCrea}(1964)}]{1964McCrea}
{McCrea}, W.~H. 1964, \mnras, 128, 147, \dodoi{10.1093/mnras/128.2.147}

\bibitem[{{McLaughlin} \& {van der Marel}(2005)}]{McLaughlin2005}
{McLaughlin}, D.~E., \& {van der Marel}, R.~P. 2005, \apjs, 161, 304,
  \dodoi{10.1086/497429}

\bibitem[{{Miocchi} {et~al.}(2013){Miocchi}, {Lanzoni}, {Ferraro},
  {Dalessandro}, {Vesperini}, {Pasquato}, {Beccari}, {Pallanca}, \&
  {Sanna}}]{miocchi2013}
{Miocchi}, P., {Lanzoni}, B., {Ferraro}, F.~R., {et~al.} 2013, \apj, 774, 151,
  \dodoi{10.1088/0004-637X/774/2/151}

\bibitem[{{Raso} {et~al.}(2019){Raso}, {Pallanca}, {Ferraro}, {Lanzoni},
  {Mucciarelli}, {Origlia}, {Dalessandro}, {Bellini}, {Libralato}, \&
  {Anderson}}]{raso2019}
{Raso}, S., {Pallanca}, C., {Ferraro}, F.~R., {et~al.} 2019, \apj, 879, 56,
  \dodoi{10.3847/1538-4357/ab2637}

\bibitem[{{Raso} {et~al.}(2020){Raso}, {Libralato}, {Bellini}, {Ferraro},
  {Lanzoni}, {Cadelano}, {Pallanca}, {Dalessandro}, {Piotto}, {Anderson}, \&
  {Sohn}}]{Raso2020}
{Raso}, S., {Libralato}, M., {Bellini}, A., {et~al.} 2020, \apj, 895, 15,
  \dodoi{10.3847/1538-4357/ab8ae7}

\bibitem[{{Sandage}(1953)}]{Sandage1953}
{Sandage}, A.~R. 1953, \aj, 58, 61, \dodoi{10.1086/106822}

\bibitem[{{Shara} {et~al.}(1997){Shara}, {Saffer}, \& {Livio}}]{shara1997}
{Shara}, M.~M., {Saffer}, R.~A., \& {Livio}, M. 1997, \apjl, 489, L59,
  \dodoi{10.1086/310952}

\bibitem[{{Song} {et~al.}(2019){Song}, {Mateo}, {Mackey}, {Olszewski},
  {Roederer}, {Walker}, \& {Bailey}}]{song2019}
{Song}, Y.-Y., {Mateo}, M., {Mackey}, A.~D., {et~al.} 2019, \mnras, 490, 385,
  \dodoi{10.1093/mnras/stz2502}

\bibitem[{{Zivick} {et~al.}(2019){Zivick}, {Kallivayalil}, {Besla}, {Sohn},
  {van der Marel}, {del Pino}, {Linden}, {Fritz}, \& {Anderson}}]{zivick2019}
{Zivick}, P., {Kallivayalil}, N., {Besla}, G., {et~al.} 2019, \apj, 874, 78,
  \dodoi{10.3847/1538-4357/ab0554}

\end{thebibliography}

\end{document}